\documentclass[a4paper,DIV11,bibtotoc]{scrartcl}
\usepackage[latin1]{inputenc}
\usepackage[T1]{fontenc}
\usepackage{array,calc,multicol,url}
\usepackage[english]{babel,varioref}
\usepackage{ae}
\usepackage[dvips,
  bookmarks,bookmarksopen,bookmarksnumbered,
  pdftitle={Proposed Specification of a Distributed XML-Query Network},
  pdfauthor={Christian Thiemann, Michael Schlenker, Thomas Severiens}
]{hyperref}

\makeatletter

\def\|#1|{\texttt{#1}}
\def\vbar{|}

\setcaphanging

\def\bnf{%
  \noindent
  \list{}{\leftmargin1em\rightmargin\leftmargin}\item\relax
  \begin{tabular}{>{\ttfamily}r>{\ttfamily}c>{\ttfamily}l}
}
\def\endbnf{
  \end{tabular}
  \endlist
}

\def\itemerr#1#2{\item[{\|[#1] #2|}]\mbox{}\\}%

\def\itemmalg#1{\item[\|#1|]\mbox{}\\}%

\newcounter{cdhgt}
\newcounter{cdpos}
\def\commdiag#1#2#3{
  \setcounter{cdhgt}{2*#1+2}
  \begin{picture}(30,\thecdhgt)(0,0)
  \put(0,0){\framebox(4,\thecdhgt){#2}}
  \put(26,0){\framebox(4,\thecdhgt){#3}}
  \setcounter{cdpos}{\thecdhgt-2}
}
\def\endcommdiag{
  \end{picture}
}
\def\msgltr#1{
  \put(4,\thecdpos){\vector(1,0){22}}
  \put(4,\thecdpos){\makebox(22,1.5){#1}}
  \setcounter{cdpos}{\thecdpos-2}
}
\def\msgrtl#1{
  \put(26,\thecdpos){\vector(-1,0){22}}
  \put(4,\thecdpos){\makebox(22,1.5){#1}}
  \setcounter{cdpos}{\thecdpos-2}
}
\def\msgmisc#1{
  \put(15,\thecdpos){\vector(1,0){11}}
  \put(15,\thecdpos){\vector(-1,0){11}}
  \put(4,\thecdpos){\makebox(22,1.5){#1}}
  \setcounter{cdpos}{\thecdpos-2}
}

\def\msglist{%
  \mbox{}\\[0.5\baselineskip]
  \begingroup\small
  \begin{tabular}{>{\ttfamily}p{\linewidth-2\tabcolsep}}%
  \hline
}
\def\endmsglist{%
  \end{tabular}\endgroup
  \par\vspace{0.5\baselineskip}%
}
\def\msg#1{#1\\\hline}
\def\crlf{$\hookleftarrow$\@ifstar{}{\newline}}

\makeatother

\begin{document}

\title{Proposed Specification of a\\Distributed XML-Query Network}
\author{Christian Thiemann \and 
        Michael Schlenker \and 
        Thomas Severiens\footnote{\|Severiens@ISN-Oldenburg.de|}
        \and {\large Institute for Science Networking Oldenburg}}
\date{October 8th 2003}
\maketitle
\begin{abstract}
  This document specifies \emph{Distributed XML-Query (DXQ)}, a method to
  query distributed XML-databases and XML-representations via a central
  interface.\vspace{0.5\baselineskip}
  
  W3C's XML-Query language~\cite{xmlquerygroup} offers a powerful instrument
  for information retrieval on XML repositories.  Here we describe an
  implementation of this retrieval in a real world's scenario.  Distributed
  XML-Query processing reduces load on every single attending node to an
  acceptable level.  The network allows every participant to control their
  computing load themselves.  Furthermore XML-repositories may stay at the
  rights holder, so every Data-Provider can decide, whether to process
  critical queries or not.  If Data-Providers keep redundant information, this
  distributed network improves reliability of information with duplicates
  removed.
\end{abstract}
\newpage
\tableofcontents
\newpage

\section{Architecture}

Every Distributed XML-Query Network (\emph{DXQ-Network}) consists of at least
one XML-Query Distributor and at least one XML-Document Provider.  The
\emph{XML-Document Provider} places an XML document~\cite{xmldoc} at the
DXQ-Network's disposal.  The \emph{XML-Query Distributor} enables the
simultaneous execution of XML-Queries~\cite{xmlquery} on all XML-repositories
in the DXQ-Network and aggregates the individual XML-Query results.  The
DXQ-Network is being used via a \emph{DXQ-Client} interface.

Any instance within the DXQ-Network such as DXQ-Client, XML-Document Provider,
and XML-Query Distributor, is a \emph{DXQ-Node}.

\subsection{XML-Document Provider}

The XML-Document Provider (\emph{XDP}) is the instance exporting its
XML-database or XML-representation into the DXQ-Network.  It offers an
interface to receive XML-Queries from the DXQ-Network to be executed on the
exported XML-document. The query result will be serialized to
XML~\cite{xqserialization} and returned to the XQD.

\subsection{XML-Query Distributor}

The XML-Query Distributor (\emph{XQD}) is the instance offering an interface
to receive XML-Queries from DXQ-Clients.  The XQD distributes the XML-Queries
to the XDPs in the DXQ-Network.  It also collects and joins the query-results.
To join the results, the XQD uses an algorithm (see section~\vref{sec:malg})
being specified by the DXQ-Client.  The XQD sends the joined results as the
result of the received XML-Query back to the DXQ-Client.

Those XDP which receive XML-Queries from the XQD are listed in the
\emph{Distribution List}.  To be listed, the XDPs have to register themselves.
The XQD also offers an interface for the registering communication with XDPs
(see~\vref{sec:dxqp-xqdxdp-control}).

\subsection{DXQ-Client}

The DXQ-Client is the instance sending XML-Queries into the DXQ-Network and
specifying algorithms for joining result sets by the XQD.

\section{Distributed XML-Query Protocol}\label{sec:dxqp}

The \emph{Distributed XML-Query Protocol (DXQP)} is the basis for every
communication between the DXQ-Nodes.  This protocol is case sensitive and
characters must\footnote{Words like 'may', 'must', 'should' are always meant
  in terms of RFC 2119.} be UTF-8~\cite{utf8} encoded.

\subsection{Transport}\label{sec:dxqp-transport}

The transportation method of the DXQP-messages between the DXQ-Nodes is
implementation depending.

\paragraph{Example HTTP:} The transport of the DXQP-messages
can be implemented by using (S)HTTP.  For this all XQDs and XDPs have to
implement their own HTTP-Server of course or be bound to an external server.
The DXQ-Client connects an XQD (or the XQD connects an XDP) by sending an
HTTP-POST-message (or GET-message, which is not recommended by the authors of
the document) to the corresponding web-server.  The message contains the
DXQP-message in its body.  Host, port and path information of the querying
DXQ-Node are derived from the Identifier of the node.  In this case the
Identifier is an URL (see also:~\vref{sec:dxqp-identname}).  The web-server
responds with the DXQP-reply-message in the body of an HTTP-message.

\paragraph{Example TCP:} TCP can also be used for transport directly by
opening separated ports bound to the nodes in the DXQ-Network.  The Identifier
of each DXQ-Node contains the port-information (e.\,g.\ 
\|dxqp://physnet.isn-oldenburg.de:8750/|).  Those of the DXQ-Nodes initiating
the communication, open channels to the port and send DXQP-messages.  The
communication accepting DXQ-Nodes respond with DXQP-messages.  The
communication channel may be kept open and can be reused for sending queries
to reduce communication overhead.

\subsection{Name and Identifier of a DXQ-Node}\label{sec:dxqp-identname}

The \emph{Name} is a string describing any DXQ-Node in a user-friendly way.
The Name should be unique within the DXQ-Network.  The Name may contain every
character apart from \|<CRLF>|, \|\{|, and \|\}|.  The \emph{Identifier} is
used for unique identification of any DXQ-Node within the DXQ-Network.
\begin{bnf}
NODE-NAME &::=& [\emph{any character sequence not containing}\\
             && \ \emph{<CRLF>, '\{' or '\}'}]\\\
NODE-IDENTIFIER &::=& (URL | "")\\
\end{bnf}

For any XQD and XDP the URL~\cite{url} of the interface of the specific
DXQ-Node is used as the Identifier (e.\,g.\ 
\|http://physnet.isn-oldenburg.de/dxq-xdp/| or
\|dxqp://physnet.isn-oldenburg.de:8750/|).  The DXQ-Client uses the empty
string as its Identifier, until an Identifier is assigned to the DXQ-Client by
the XQD.  The assigned Identifier is passed within the header variable
\|Msg-To| of the first message transferred from the XQD to the DXQ-Client as a
reply to the empty string in the header variable \|Msg-From| in the former
message (see~also:~\vref{sec:dxqp-msg-structure}).  The Identifier has to be a
valid URL and unique for the DXQ-Client and the XQD.  The following examples
are valid Identifiers: \|http://a6bf278d|, \|http://134.106.31.210/ab6bf278d|
or \|http://134.106.31.210/?sessid=a6bf278d|.

\subsection{Structure of a DXQP-Message}\label{sec:dxqp-msg-structure}

Every \emph{DXQP-message} consists of a header and a body.  The two parts are
separated by \|<CRLF>|.  The header consists of the message's \|ID-LINE| and
several \emph{header variables}.  The body of the message has to contain the
number of bytes given in the header variable \|Content-Length|.  If
\|Content-Length| is empty or not a positive integer, the message will end
with the \|<CRLF>| closing the header part.

\newpage The general grammar for any DXQP-message is:
\begin{bnf}
MESSAGE &::=& HEADER <CRLF> BODY\\
HEADER &::=& ID-LINE <CRLF> (VARIABLE)*\\
ID-LINE &::=& "DXQP-" VERSION " " MSGTYPE\\
VERSION &::=& [0-9] "." [0-9]\\
MSGTYPE &::=& ("OK" | "ERROR"|\\
           && \ "XML-QUERY" | "MERGE-ALGORITHM" |\\
           && \ "XML-QUERY-RESULT" | "XML-QUERY-MERGED-RESULT" |\\
           && \ "REGISTER" | "UNREGISTER" |\\
           && \ "ADDTODL" | "RMFROMDL" |\\
           && \ "INFO-REQUEST" | "INFO-REPLY")\\
VARIABLE &::=& VNAME ":" (" ")+ VVALUE <CRLF>\\
VNAME &::=& [A-Za-z-]+\\
VVALUE &::=& [\emph{any byte sequence not containing <CRLF>}]\\
<CRLF> &::=& 0x0D 0x0A\\
BODY &::=& [\emph{any byte sequence}]\\
\end{bnf}

The number, meaning and content of header variables in a message are
implementation depending.  Header variables in this document are given as
examples only, but implementations should not use variables with identical
names but different semantics.  This applies to the already defined variable
\|Content-Length| as well as to \|Msg-From| and \|Msg-To|.

The header variable \|Msg-From| gives the Identifier of the DXQ-Node sending
the message.  \|Msg-To| gives the Identifier of the DXQ-Node receiving the
message.  Without a valid Identifier (see~also:~\vref{sec:dxqp-identname}) in
any of these variables the complete message becomes invalid.
\begin{bnf}
VAR-Msg-From &::=& "Msg-From:\ " NODE-IDENTIFIER <CRLF>\\
VAR-Msg-To &::=& "Msg-To:\ " NODE-IDENTIFIER <CRLF>\\
VAR-Content-Length &::=& "Content-Length:\ " [0-9]+ <CRLF>\\
\end{bnf}

\subsection{DXQP-1.0 Messages}

\subsubsection{\|OK|}\label{msg:OK}
The \|OK|-message gives a positive acknowledgment.
\begin{bnf}
MSG-OK &::=& "DXQP-1.0 OK" <CRLF>\\
          && VAR-Msg-From VAR-Msg-To\\
          && (VAR-Transaction-ID)?\\
          && <CRLF>\\
\end{bnf}
The header variable \|Transaction-ID| is exclusive and mandatory for responses
on \|XML-QUERY|-messages only (see~also:~\vref{msg:XML-QUERY} and
section~\vref{sec:dxqp-comm-clientxqd}).

\subsubsection{\|ERROR|}\label{msg:ERROR}
The \|ERROR|-message gives a negative acknowledgment and reports of errors.
This message has to comprise the header variable \|Error-Code| giving the
value of the error-code as defined in \vref{sec:dxqp-1.0-errorcodes}.  The
body of the message may contain further error information.
\begin{bnf}
MSG-ERROR &::=& "DXQP-1.0 ERROR" <CRLF>\\
             && VAR-Msg-From VAR-Msg-To\\
             && VAR-Error-Code\\
             && (<CRLF> | (VAR-Content-Length <CRLF> BODY))\\
VAR-Error-Code &::=& "Error-Code:\ " ERROR-CODE <CRLF>\\
ERROR-CODE &::=& [0-9]\{3\}\\
\end{bnf}

\subsubsection{\|XML-QUERY|}\label{msg:XML-QUERY}
The \|XML-QUERY|-message provides an XML-Query in its body for execution by
the receiving DXQ-Node.  If this message is sent to an XQD it has to comprise
the header variable \|Merge-Algorithm|.\par The header variable
\|Transaction-ID| gives a \|VVALUE| arbitrarily chosen by the sender of the
message to allow unique identification of this message in the context of
multi-query processing.  The \|Transaction-ID| must not contain any white
spaces.
\begin{bnf}
MSG-XML-QUERY &::=& "DXQP-1.0 XML-QUERY" <CRLF>\\
                 && VAR-Msg-From VAR-Msg-To\\
                 && VAR-Transaction-ID (VAR-Merge-Algorithm)?\\
                 && VAR-Content-Length <CRLF> XML-QUERY\\
VAR-Transaction-ID &::=& "Transaction-ID:\ " TRANSACTION-ID <CRLF>\\
TRANSACTION-ID &::=& [\emph{any VVALUE not containing " "}]\\
VAR-Merge-Algorithm &::=& "Merge-Algorithm:\ " MALG-SPEC <CRLF>\\
MALG-SPEC &::=& [a-z0-9-]\\
\end{bnf}

\subsubsection{\|MERGE-ALGORITHM|}\label{msg:MERGE-ALGORITHM}
After the DXQ-Client sent an \|XML-QUERY|-message to the XQD giving
\|"user-defined"| in \|Merge-Algorithm|, it has to send the
\|MERGE-ALGORITHM|-message, giving an XML-Query in its body to join the
query-results of the single XDP in the DXQ-Network.\par The header variable
\|Transaction-ID| has to provide the same value as given in the corresponding
\|XML-QUERY|-message.
\begin{bnf}
MSG-MERGE-ALGORITHM &::=& "DXQP-1.0 MERGE-ALGORITHM" <CRLF>\\
                       && VAR-Msg-From VAR-Msg-To\\
                       && VAR-Transaction-ID\\
                       && VAR-Content-Length <CRLF> XML-QUERY
\end{bnf}

\subsubsection{\|XML-QUERY-RESULT|}\label{msg:XML-QUERY-RESULT}
The \|XML-QUERY-RESULT|-message is the answer to an \|XML-QUERY|-message.  It
is sent from every XDP to the XQD. The serialized query result is given in the
body of this message.  The header variable \|Transaction-ID| has to give the
same value as given in the corresponding \|XML-Query|-message.
\begin{bnf}
MSG-XML-QUERY-RESULT &::=& "DXQP-1.0 XML-QUERY-RESULT" <CRLF>\\
                        && VAR-Msg-From VAR-Msg-To\\
                        && VAR-Transaction-ID\\
                        && VAR-Content-Length <CRLF> XML-DOCUMENT\\
\end{bnf}

\subsubsection{\|XML-QUERY-MERGED-RESULT|}\label{msg:XML-QUERY-MERGED-RESULT}
The \|XML-QUERY-MERGED-RESULT|-message is the answer to the
\|XML-QUERY|-message.  It is sent from the XQD to the DXQ-Client. The
serialized joined result of the corresponding query is given in the body of
the message.  The header variable \|Transaction-ID| has to give the same value
as given in the corresponding \|XML-Query|-message.  The header variable
\|Result-Sources| should contain a list of Names of those XDPs which delivered
parts of the result set.
\begin{bnf}
\multicolumn{3}{l}{\|MSG-XML-QUERY-MERGED-RESULT\hspace{2\tabcolsep}::=|}\\
 && "DXQP-1.0 XML-QUERY-MERGED-RESULT" <CRLF>\\
 && VAR-Msg-From VAR-Msg-To\\
 && VAR-Transaction-ID VAR-Result-Sources\\
 && VAR-Content-Length <CRLF> XML-DOCUMENT\\
\multicolumn{3}{l}{\|VAR-Result-Sources\hspace{2\tabcolsep}::=|}\\
 && "Result-Sources:\ " ("\{" NODE-NAME "\} ")*\\
 && \mbox{}~~~~~~~~~~~~~~~~~~~~~"\{" NODE-NAME "\}" <CRLF>\\
\end{bnf}

\subsubsection{\|REGISTER|}\label{msg:REGISTER}
By sending the \|REGISTER|-message, the XDP registers itself at the XQD.
\begin{bnf}
MSG-REGISTER &::=& "DXQP-1.0 REGISTER" <CRLF>\\
                && VAR-Msg-From VAR-Msg-To\\
                && VAR-Node-Name\\
                && <CRLF>\\
VAR-Node-Name &::=& "Node-Name:\ " NODE-NAME <CRLF>\\
\end{bnf}

\subsubsection{\|UNREGISTER|}\label{msg:UNREGISTER}
By sending the \|UNREGISTER|-message, the XDP checks out from the XQD.
\begin{bnf}
MSG-UNREGISTER &::=& "DXQP-1.0 UNREGISTER" <CRLF>\\
                  && VAR-Msg-From VAR-Msg-To\\
                  && <CRLF>\\
\end{bnf}

\subsubsection{\|ADDTODL|}\label{msg:ADDTODL}
The \|ADDTODL|-message is used by the XDP to sign in to the Distribution List
of the XQD.  The XQD will distribute queries to those XDPs listed in the
Distribution List only.
\begin{bnf}
MSG-ADDTODL &::=& "DXQP-1.0 ADDTODL" <CRLF>\\
               && VAR-Msg-From VAR-Msg-To\\
               && <CRLF>\\
\end{bnf}

\subsubsection{\|RMFROMDL|}\label{msg:RMFROMDL}
To sign off the Distribution List of the XQD, the XDP sends the
\|RMFROMDL|-message.
\begin{bnf}
MSG-RMFROMDL &::=& "DXQP-1.0 RMFROMDL" <CRLF>\\
                && VAR-Msg-From VAR-Msg-To\\
                && <CRLF>\\
\end{bnf}

\subsubsection{\|INFO-REQUEST|}\label{msg:INFO-REQUEST}
To get information about an XQD or XDP, every DXQ-Node can send the
\|INFO-REQUEST|-message.  The header variable \|Request| contains a list of
inquired information.  To inquire a sign of life only, the header variable
\|Request| may be empty.  \|INFO-NAME| is described in \vref{msg:INFO-REPLY}
in more detail.  Instead of the detailed list of information, the \|Request|
variable can also contain \|"*"| to ask for all information available.
\begin{bnf}
MSG-INFO-REQUEST &::=& "DXQP-1.0 INFO-REQUEST" <CRLF>\\
&& VAR-Msg-From VAR-Msg-To\\
&& VAR-Request\\
&& <CRLF>\\
VAR-Request &::=& "Request:\ " \\
&& ("*" | (INFO-NAME " ")* INFO-NAME)? <CRLF>\\
\end{bnf}

\subsubsection{\|INFO-REPLY|}\label{msg:INFO-REPLY}
The \|INFO-REPLY|-message is the answer to the \|INFO-REQUEST|-message.  This
message contains all of the requested information in the corresponding header
variables.
\begin{bnf}
MSG-INFO-REPLY &::=& "DXQP-1.0 INFO-REPLY" <CRLF>\\
                  && VAR-Msg-From VAR-Msg-To\\
                  && (VAR-INFO)*\\
                  && <CRLF>\\
VAR-INFO &::=& INFO-NAME ":\ " INFO-VALUE <CRLF>\\
INFO-NAME &::=& VNAME\\
INFO-VALUE &::=& VVALUE\\
YESNO &::=& ("yes" | "no")\\
MALG-LIST &::=& ((MALG-SPEC " ")* MALG-SPEC)?\\
XDP-LIST &::=& ((XDP-SPEC " ")* XDP-SPEC)?\\
XDP-SPEC &::=& (NODE-IDENTIFIER)? "\{" NODE-NAME "\}"\\
TRANSACTION-ID-LIST &::=& ((TRANSACTION-ID " ")* TRANSACTION-ID)?\\
\end{bnf}\par
Meaning and Values of \|INFO-NAME| are implementation depending. Several
\|INFO-NAME|s are pre-defined with this specification.  If an \|INFO-REQUEST|
asks for an \|INFO-NAME| which is not supported by the specific
implementation, an empty value, not an error will be returned in
\|INFO-REPLY|.

Here several \|INFO-NAME|s are pre-defined, in parentheses the name of the
grammar-pro\-duction of the value in the \|INFO-REPLY|-message is given.
\def\infodesc#1#2#3{\item[\|#1| (\|#2|)]\mbox{}\\{#3}}%
\begin{description}
\infodesc{Node-Name}{NODE-NAME}
  {Name of the DXQ-Node (see~also:~\vref{sec:dxqp-identname}).}
\infodesc{Admin}{VVALUE}
  {Information on the administrator of the DXQ-Node (name, email, etc.)}
\infodesc{Registered}{YESNO}
  {The value of this variable will be \|"yes"|, if it is asked by an XDP
  to an XQD and the XDP is registered at the XQD. Else the value will be
  \|"no"|.}
\infodesc{Is-in-DL}{YESNO}
  {The value of this variable will be \|"yes"|, if it is asked by an XDP
  to an XQD and the XDP is listed in the Distribution List of the XQD. Else
  the value will be \|"no"|.}
\infodesc{Merge-Algorithms}{MALG-LIST}
  {Delivers a list of implemented merging algorithms (including the mandatory
  algorithm \|user-defined|). If the DXQ-Node to be asked is not an XQD, the
  value delivered by this variable will be the empty string.}
\infodesc{Registered-XDPs}{XDP-LIST}
  {Delivers a list of registered XDPs. It is optional to support this
  information.}
\infodesc{Active-XDPs}{XDP-LIST}
  {Delivers the list of XDPs from the Distribution List. It is optional to
  support this information.}
\infodesc{Active-Queries}{TRANSACTION-ID-LIST}
  {Delivers a list of the Transaction-IDs initiated by the DXQ-Node being
  asking and still in progress at the DXQ-Node to be asked. The list must only
  contain those Transaction-IDs processed under the same Identifier as those
  of the \|INFO-REQUEST|-message. It is optional to support this information.}
\end{description}

\subsection{DXQP-1.0 Error-Codes}\label{sec:dxqp-1.0-errorcodes}

\begin{description}
\itemerr{100}{Invalid message}
  The \|ID-LINE|, the name of one of the header variables or the Identifier in
  the header variable \|Msg-From| or \|Msg-To| is invalid.
\itemerr{101}{Unexpected message}
  The message was not expected in the current context (e.\,g.\ \|REGISTER|
  sent to an XDP), although it has the correct syntax.
\itemerr{102}{Missing header variable}
  The message lacks of at least one header variable, essential for the current
  context of the message. The body of this message contains the name of the
  missing variable.
\itemerr{103}{Missing content}
  Though it is essential for processing, the body part of the message is
  missing (e.\,g.\ in an \|XML-Query|-message).
\itemerr{200}{XML-Query processor error}
  The XML-Query processor/interpreter gave an error. The body of this
  message contains the original error-message from the
  processor/interpreter.
\itemerr{300}{Unsupported merge algorithm}
  The XQD does not support the merging algorithm specified in the header
  variable \|Merge-Algorithm| of an \|XML-QUERY|-message.
\itemerr{400}{No XML document providers available}
  If the Distribution List of an XQD is empty, this \|ERROR|-message will be
  replied on \|XML-QUERY|-messages.
\itemerr{500}{Internal error}
  An internal error occurred when processing the message.
\itemerr{9xx}{Implementation-defined error}
  All \|9xx| error codes are reserved for implementation defined error-codes.
\end{description}

\subsection{Communication}

Here the semantics of the DXQP-messages within several contexts are
described.

\subsubsection{XQD and XDP (Control)}\label{sec:dxqp-xqdxdp-control}

\paragraph{Register:} 
Communication is initiated by sending a \|REGISTER|-message from the XDP to
the XQD. If the XDP was registered at the XQD, the XQD replies the
\|OK|-message, else the \|ERROR|-message will be replied.

When sending the \|UNREGISTER|-message, the XQD removes the corresponding XDP
from the list of registered XDPs and replies with the \|OK|-message. In case
of an error during processing the \|UNREGISTER|-message, the \|ERROR|-message
will be sent. To leave the undefined registration status, the XDP may use the
\|INFO-REQUEST|-message.

The period between \|REGISTER|--\|OK| and \|UNREGISTER|--\|OK| is called a
\emph{Session}.

\begin{figure}[htbp]
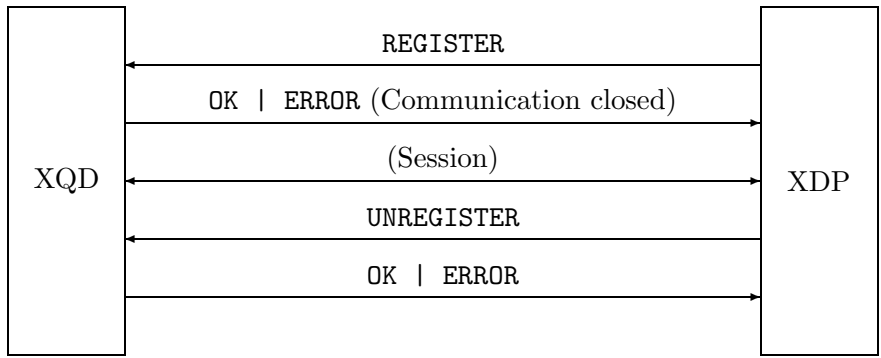
\centering
  \begin{commdiag}{5}{XQD}{XDP}
    \msgrtl{\|REGISTER|}
    \msgltr{\|OK \vbar\ ERROR| (Communication closed)}
    \msgmisc{(Session)}
    \msgrtl{\|UNREGISTER|}
    \msgltr{\|OK \vbar\ ERROR|}
  \end{commdiag}%
  \caption{\|REGISTER|-/\|UNREGISTER|-Communication}%
  \label{fig:com-register}%
\end{figure}

\paragraph{Distribution List:}
Every XQD runs a list of registered XDPs, which will get XML-Queries in the
distribution process.  XDPs have to sign themselves into this
\emph{Distribution List} explicitly.

To sign into the Distribution List, the XDP sends the \|ADDTODL|-message to
the XQD. The XQD will answer either with the \|OK|- or with the
\|ERROR|-message.

To verify the status of the Distribution List, the XDP should use the
\|INFO-REQUEST|-message frequently (see~also:~\vref{msg:INFO-REQUEST}).

The XDP may sign off from the Distribution List (e.\,g.\ to reduce the server
load in critical situations) by sending the \|RMFROMDL|-message to the XQD.
The XQD will reply with an \|OK|-message. In case of an error the
\|ERROR|-message will be sent. To request its own status in the Distribution
List, the XDP should use the \|INFO-REQUEST|-message.

\begin{figure}[htbp]
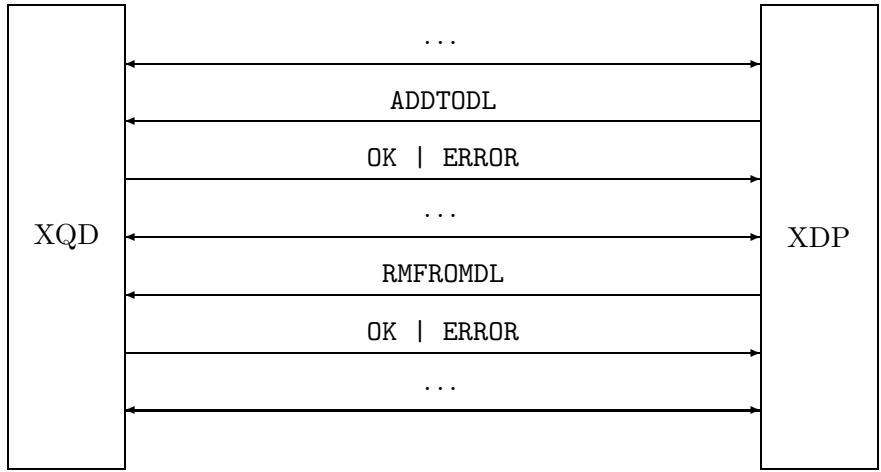
\centering
  \begin{commdiag}{7}{XQD}{XDP}
    \msgmisc{\ldots}
    \msgrtl{\|ADDTODL|}
    \msgltr{\|OK \vbar\ ERROR|}
    \msgmisc{\ldots}
    \msgrtl{\|RMFROMDL|}
    \msgltr{\|OK \vbar\ ERROR|}
    \msgmisc{\ldots}
  \end{commdiag}%
  \caption{Sign into and sign off the Distribution List}%
  \label{fig:dl}%
\end{figure}

\paragraph{Connectivity Care:}
Every XQD should send \|INFO-REQUEST|-messages with empty \|Request|-variable
to all registered XDP in regular intervals. Every XDP has to reply with an
\|INFO-REPLY|-message. In case of no reply message, a time-out, or an
\|ERROR|-reply, the XQD should remove the corresponding XDP from the
Distribution List.

After removing an XDP from the Distribution List, an XQD may still send
\|INFO-REQUEST|-messages to the still registered XDPs. In the case of repeated
time-outs the XQD may unregister the XDP.

Every XDP may send \|INFO-REQUEST|-messages to verify being registered and
signed into the Distribution List.  After being removed from the Distribution
List or unregistered, the XDP may register and sign in again.

\begin{figure}[htbp]
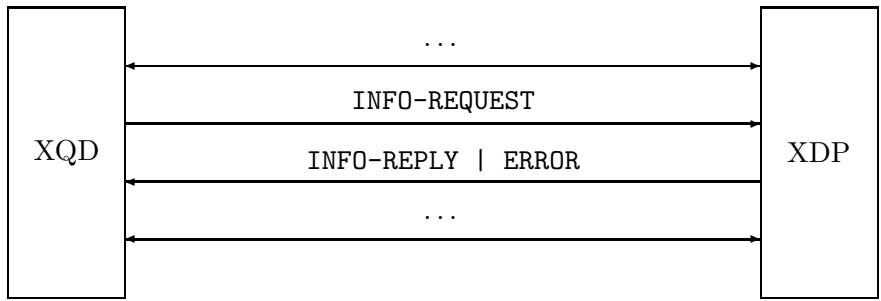
\centering
  \begin{commdiag}{4}{XQD}{XDP}
    \msgmisc{\ldots}
    \msgltr{\|INFO-REQUEST|}
    \msgrtl{\|INFO-REPLY \vbar\ ERROR|}
    \msgmisc{\ldots}
  \end{commdiag}%
  \caption{Ping via \|INFO-REQUEST| and \|INFO-REPLY|}%
  \label{fig:ping}%
\end{figure}

\subsubsection{XQD and XDP (Data)}
Every XDP listed in the Distribution List of an XQD gets \|XML-QUERY|-messages
which contain the XML-Queries to be processed on the XML-Document,
XML-Database or XML-Representation.  The XDP will reply an
\|XML-QUERY-RESULT|- or an \|ERROR|-message.

If the XML-Query succeeds, the \|XML-QUERY-RESULT|-message will reply in its
body part the result of the XML-Query executed.  In case of any error during
the processing of the XML-Query, the \|ERROR|-message will be replied giving
the respective \|Error-Code|.  The error-message of the XML-Query processor
should be given in the body part of the \|ERROR|-message.

\begin{figure}[htbp]
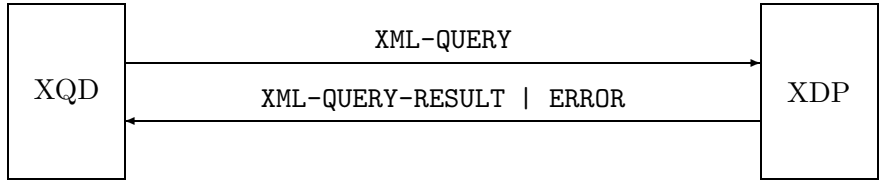
\centering
  \begin{commdiag}{2}{XQD}{XDP}
    \msgltr{\|XML-QUERY|}
    \msgrtl{\|XML-QUERY-RESULT \vbar\ ERROR|}
  \end{commdiag}%
  \caption{Datatransfer between XQD and XDP}%
  \label{fig:xml-query-xqdxdp}%
\end{figure}

\subsubsection{DXQ-Client and XQD (Data)}\label{sec:dxqp-comm-clientxqd}
The DXQ-Client (user) initiates the communication by sending the
\|XML-QUERY|-message to the XQD.  This \|XML-QUERY|-message contains the
XML-Query to be processed by the XDP in its body. The further communication
depends on the value of the header variable \|Merge-Algorithm| of this
message.

If the value of the header variable \|Merge-Algorithm| is \|user-defined|,
then the XQD will reply an \|OK|-message and will wait for the
\|MERGE-ALGORITHM|-message to be sent by the DXQ-Client.  All messages of this
query handling (\|XML-QUERY|-message, \|OK|-message,
\|MERGE-ALGORITHM|-message) have to give the same value in the
\|Transaction-ID| header variable, to allow unique assignment of the messages.

If the value of the header variable \|Merge-Algorithm| is not \|user-defined|,
then the XQD will reply the \|XML-QUERY-MERGED-RESULT|-message directly,
joined with the selected of the pre-defined algorithms (see also:
section~\vref{sec:malg}).

In the case of any error while processing the XML-Query, the XQD sends the
\|ERROR|-message, which closes the communication.

The DXQ-Client sends the \|MERGE-ALGORITHM|-message with the XML-Query in its
body to join the result sets.  For further information see also
section~\vref{sec:malg}.  With the \|OK|-message message the DXQ assigns an
Identifier to the DXQ-Client within the \|Msg-To| header variable. This
Identifier has to be used in the further communication, for example in the
\|Msg-From| header variable of the \|MERGE-ALGORITHM|-message.

The XQD will reply the \|XML-QUERY-MERGED-RESULT|-message, which contains the
joined result in its body. In the case of an error, the XQD replies the
\|ERROR|-message. If only some but not all of the XDP reply \|ERROR|-messages,
the XQD should process the available results.  The header variable
\|Result-Sources| lists those XDPs only which delivered data for the joined
result. In case of all XDPs replying an \|ERROR|-message, the XQD will also
reply an \|ERROR|-message.

\begin{figure}[htbp]
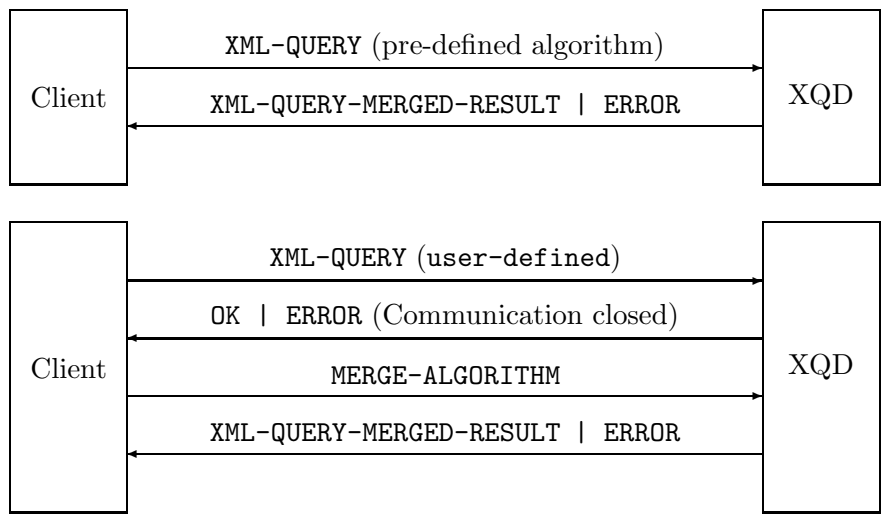
\centering
  \begin{commdiag}{2}{Client}{XQD}
    \msgltr{\|XML-QUERY| (pre-defined algorithm)}
    \msgrtl{\|XML-QUERY-MERGED-RESULT \vbar\ ERROR|}
  \end{commdiag}\\[\baselineskip]
  \begin{commdiag}{4}{Client}{XQD}
    \msgltr{\|XML-QUERY| (\|user-defined|)}
    \msgrtl{\|OK \vbar\ ERROR| (Communication closed)}
    \msgltr{\|MERGE-ALGORITHM|}
    \msgrtl{\|XML-QUERY-MERGED-RESULT \vbar\ ERROR|}
  \end{commdiag}%
  \caption{Data-transfer between DXQ-Client and XQD}%
  \label{fig:xml-query-clientxqd}%
\end{figure}

\section{Merge-Algorithm}\label{sec:malg}

To join the different results of the single XDPs, the XQD uses a
\emph{Merge-Algorithm} which is specified by the DXQ-Client in the
\|XML-QUERY|-message (see also:~\vref{msg:XML-QUERY}).  The DXQ-Client can ask
the XQD for a list of available pre-defined merging algorithms by sending an
\|INFO-REQUEST|-message (see also:~\vref{msg:INFO-REQUEST}). The resulting
list of algorithms is implementation depending.

Here two Merge-Algorithms are described in detail.  While the
\|concatenate|-algorithm is mandatory for every XQD implementation, the
\|remove-duplicates|-algorithm is an example of an implementation depending,
non-obligatory algorithm.

\begin{description}
\itemmalg{concatenate}
  The results of the XDPs are just sticked together and framed by a
  \|<result>|-Tag.
\itemmalg{remove-duplicates}
  Assuming the following pair of results was delivered by two XDPs:
  {\footnotesize
  \begin{multicols}{2}
  \|\mbox{}<solarsystem>\\
    \mbox{}~~<planets>\\
    \mbox{}~~~~<planet>Mercury</planet>\\
    \mbox{}~~~~<planet>Venus</planet>\\
    \mbox{}~~~~<planet>Earth</planet>\\
    \mbox{}~~</planets>\\
    \mbox{}</solarsystem>|\\
  \|\mbox{}<solarsystem>\\
    \mbox{}~~<planets>\\
    \mbox{}~~~~<planet>Venus</planet>\\
    \mbox{}~~~~<planet>Earth</planet>\\
    \mbox{}~~~~<planet>Mars</planet>\\
    \mbox{}~~</planets>\\
    \mbox{}</solarsystem>|
  \end{multicols}}
  The \|remove-duplicates|-algorithm will join these results into:
  {\footnotesize
  \begin{quote}
  \|\mbox{}<solarsystem>\\
    \mbox{}~~<planets>\\
    \mbox{}~~~~<planet>Mercury</planet>\\
    \mbox{}~~~~<planet>Venus</planet>\\
    \mbox{}~~~~<planet>Earth</planet>\\
    \mbox{}~~~~<planet>Mars</planet>\\
    \mbox{}~~</planets>\\
    \mbox{}</solarsystem>|
  \end{quote}}
  The XML-document is copied down to a specified depth. All elements of this
  or higher depth are sticked together, duplicates are removed. To specify
  this critical depth, this algorithm reads the header-variable \|Depth|.
\end{description}

Every XQD has to support a mandatory \|user-defined| algorithm.  If selecting
this algorithm, the DXQ-Client must send a \|MERGE-ALGORITHM|-message to the
XQD. This message contains an XML-Query which is able to join the results
received from the XDPs together. (For detailed information on the
communication between DXQ-Client and XQD see
also:~\vref{sec:dxqp-comm-clientxqd}.)

The XQD lists all XDP-results into the context-item to be processed with the
user-defined merge-algorithm:
\begin{quote}
\|\mbox{}<context-item>\\
  \mbox{}~~<result>\\
  \mbox{}~~~~<xdp>\\
  \mbox{}~~~~~~<name>\emph{XDP-Name}</name>\\
  \mbox{}~~~~</xdp>\\
  \mbox{}~~~~<xqres>\emph{XML-Query Result}</xqres>\\
  \mbox{}~~</result>\\
  \mbox{}~~\ldots\\
  \mbox{}</context-item>|
\end{quote}
The \|<xdp>|-Tag can also contain further implementation depending
information. \emph{XML-Query Result} is the body of the
\|XML-QUERY-RESULT|-message as received from the specific XDP.

The result of the merging process will be sent to the DXQ-Client within the
XQD's \|XML-QUERY-MERGED-RESULT|-message.

\appendix
\section{Changes}
\textbf{August 25th 2003:}
\begin{itemize}
\item To prevent misunderstandings about the header variable \|Query-ID| it
  has been renamed to \|Transaction-ID|.
\end{itemize}
\textbf{August 29th 2003:}
\begin{itemize}
\item To protect the privacy of the XDPs the \|<ident>| tag in the context
  item of the \|user-defined| merge algorithm has been removed.
  Implementations may reinsert this tag if they want to publish the XDPs'
  Indentifiers.
\end{itemize}
\textbf{October 8th 2003:}
\begin{itemize}
\item The information \|Active-Queries| (in
  \|INFO-REQUEST|/\|INFO-REPLY|-messages) has been made optional.
\item The \|NODE-IDENTIFIER| in the grammar production \|XDP-SPEC| (used for
  the information \|Registered-XDPs| and \|Active-XDPs|) has been made
  optional.
\item A new header variable \|Node-Name| has been inserted into the
  \|REGISTER|-message passing the XDP's Name to the XQD.
\end{itemize}

\newpage
\section{Example: DXQP-Communication}

Here the communication within an example DXQ-Network is described.  The
DXQ-Network consists of one XDQ and two XDPs, both exporting the XML-document
\begin{quotation}\|<document><a>5</a></document>|\end{quotation}
One DXQ-Client builds the interface of the DXQ-Network.

First both XDPs register themselves at the XQD, being included into the
Distribution List of the XQD:\footnote{\crlf* = \|<CRLF>|}
\begin{msglist}
\msg{DXQP-1.0 REGISTER\crlf
     Msg-From:\ http://physnet.isn-oldenburg.de/dxq-xdp/\crlf
     Msg-To:\ http://metasearch.isn-oldenburg.de/dxq-xqd/\crlf
     Node-Name:\ PhysNet\crlf
     \crlf*}
\msg{DXQP-1.0 OK\crlf
     Msg-From:\ http://metasearch.isn-oldenburg.de/dxq-xqd/\crlf
     Msg-To:\ http://physnet.isn-oldenburg.de/dxq-xdp/\crlf
     \crlf*}
\msg{DXQP-1.0 ADDTODL\crlf
     Msg-From:\ http://physnet.isn-oldenburg.de/dxq-xdp/\crlf
     Msg-To:\ http://metasearch.isn-oldenburg.de/dxq-xqd/\crlf
     \crlf*}
\msg{DXQP-1.0 OK\crlf
     Msg-From:\ http://metasearch.isn-oldenburg.de/dxq-xqd/\crlf
     Msg-To:\ http://physnet.isn-oldenburg.de/dxq-xdp/\crlf
     \crlf*}
\end{msglist}
\begin{msglist}
\msg{DXQP-1.0 REGISTER\crlf
     Msg-From:\ http://physnet-mirror.isn-oldenburg.de:8080/dxq-xdp/\crlf
     Msg-To:\ http://metasearch.isn-oldenburg.de/dxq-xqd/\crlf
     Node-Name:\ PhysNet (Mirror)\crlf
     \crlf*}
\msg{DXQP-1.0 OK\crlf
     Msg-From:\ http://metasearch.isn-oldenburg.de/dxq-xqd/\crlf
     Msg-To:\ http://physnet-mirror.isn-oldenburg.de:8080/dxq-xdp/\crlf
     \crlf*}
\msg{DXQP-1.0 ADDTODL\crlf
     Msg-From:\ http://physnet-mirror.isn-oldenburg.de:8080/dxq-xdp/\crlf
     Msg-To:\ http://metasearch.isn-oldenburg.de/dxq-xqd/\crlf
     \crlf*}
\msg{DXQP-1.0 OK\crlf
     Msg-From:\ http://metasearch.isn-oldenburg.de/dxq-xqd/\crlf
     Msg-To:\ http://physnet-mirror.isn-oldenburg.de:8080/dxq-xdp/\crlf
     \crlf*}
\end{msglist}

\newpage Following, the XQD inquires information about the XDPs:
\begin{msglist}
\msg{DXQP-1.0 INFO-REQUEST\crlf
     Msg-From:\ http://metasearch.isn-oldenburg.de/dxq-xqd/\crlf
     Msg-To:\ http://physnet.isn-oldenburg.de/dxq-xdp/\crlf
     Request:\ Node-Name Admin\crlf
     \crlf*}
\msg{DXQP-1.0 INFO-REPLY\crlf
     Msg-From:\ http://physnet.isn-oldenburg.de/dxq-xdp/\crlf
     Msg-To:\ http://metasearch.isn-oldenburg.de/dxq-xqd/\crlf
     Node-Name:\ PhysNet\crlf
     Admin:\ Max Mustermann <admin@physnet.isn-oldenburg.de>\crlf
     \crlf*}
\end{msglist}
\begin{msglist}
\msg{DXQP-1.0 INFO-REQUEST\crlf
     Msg-From:\ http://metasearch.isn-oldenburg.de/dxq-xqd/\crlf
     Msg-To:\ http://physnet-mirror.isn-oldenburg.de:8080/dxq-xdp/\crlf
     Request:\ Node-Name Admin\crlf
     \crlf*}
\msg{DXQP-1.0 INFO-REPLY\crlf
     Msg-From:\ http://physnet-mirror.isn-oldenburg.de:8080/dxq-xdp/\crlf
     Msg-To:\ http://metasearch.isn-oldenburg.de/dxq-xqd/\crlf
     Node-Name:\ PhysNet (Mirror)\crlf
     Admin:\ Bert Beispiel <admin@physnet-mirror.isn-oldenburg.de>\crlf
     \crlf*}
\end{msglist}

The DXQ-Client connects the XQD by sending an initial XML-Query:
\begin{msglist}
\msg{DXQP-1.0 XML-QUERY\crlf
     Msg-From:\ \crlf
     Msg-To:\ http://metasearch.isn-oldenburg.de/dxq-xqd/\crlf
     Transaction-ID:\ 0\crlf
     Merge-Algorithm:\ user-defined\crlf
     Content-Length:\ 23\crlf
     \crlf
     let \$a := ./a return \$a}
\msg{DXQP-1.0 OK\crlf
     Msg-From:\ http://metasearch.isn-oldenburg.de/dxq-xqd/\crlf
     Msg-To:\ http://a6bf\crlf
     Transaction-ID:\ 0\crlf
     \crlf*}
\end{msglist}

\newpage The XQD distributes the XML-Query to all XDPs which are listed in the
Distribution List. It also collects the results:
\begin{msglist}
\msg{DXQP-1.0 XML-QUERY\crlf
     Msg-From:\ http://metasearch.isn-oldenburg.de/dxq-xqd/\crlf
     Msg-To:\ http://physnet.isn-oldenburg.de/dxq-xdp/\crlf
     Transaction-ID:\ 0\crlf
     Content-Length:\ 23\crlf
     \crlf
     let \$a := ./a return \$a}
\msg{DXQP-1.0 XML-QUERY-RESULT\crlf
     Msg-From:\ http://physnet.isn-oldenburg.de/dxq-xdp/\crlf
     Msg-To:\ http://metasearch.isn-oldenburg.de/dxq-xqd/\crlf
     Transaction-ID:\ 0\crlf
     Content-Length:\ 8\crlf
     \crlf
     <a>5</a>}
\end{msglist}
\begin{msglist}
\msg{DXQP-1.0 XML-QUERY\crlf
     Msg-From:\ http://metasearch.isn-oldenburg.de/dxq-xqd/\crlf
     Msg-To:\ http://physnet-mirror.isn-oldenburg.de:8080/dxq-xdp/\crlf
     Transaction-ID:\ 1\crlf
     Content-Length:\ 23\crlf
     \crlf
     let \$a := ./a return \$a}
\msg{DXQP-1.0 XML-QUERY-RESULT\crlf
     Msg-From:\ http://physnet-mirror.isn-oldenburg.de:8080/dxq-xdp/\crlf
     Msg-To:\ http://metasearch.isn-oldenburg.de/dxq-xqd/\crlf
     Transaction-ID:\ 1\crlf
     Content-Length:\ 8\crlf
     \crlf
     <a>5</a>}
\end{msglist}

The XQD receives the \|MERGE-ALGORITHM|-message from the DXQ-Client
and responds with the joined results:
\begin{msglist}
\msg{DXQP-1.0 MERGE-ALGORITHM\crlf
     Msg-From:\ http://a6bf\crlf
     Msg-To:\ http://metasearch.isn-oldenburg.de/dxq-xqd/\crlf
     Transaction-ID:\ 0\crlf
     Content-Length:\ 51\crlf
     \crlf
     let \$r := <a>\{sum(./result/xqres/a)\}</a> return \$r}
\msg{DXQP-1.0 XML-QUERY-MERGED-RESULT\crlf
     Msg-From:\ http://metasearch.isn-oldenburg.de/dxq-xqd/\crlf
     Msg-To:\ http://a6bf\crlf
     Transaction-ID:\ 0\crlf
     Result-Sources:\ \{PhysNet\} \{PhysNet (Mirror)\}\crlf
     Content-Length:\ 9\crlf
     \crlf
     <a>10</a>}
\end{msglist}

\newpage Finally the XDPs send an \|UNREGISTER|-message to be removed from the
network:
\begin{msglist}
\msg{DXQP-1.0 UNREGISTER\crlf
     Msg-From:\ http://physnet.isn-oldenburg.de/dxq-xdp/\crlf
     Msg-To:\ http://metasearch.isn-oldenburg.de/dxq-xqd/\crlf
     \crlf*}
\msg{DXQP-1.0 OK\crlf
     Msg-From:\ http://metasearch.isn-oldenburg.de/dxq-xqd/\crlf
     Msg-To:\ http://physnet.isn-oldenburg.de/dxq-xdp/\crlf
     \crlf*}
\end{msglist}
\begin{msglist}
\msg{DXQP-1.0 UNREGISTER\crlf
     Msg-From:\ http://physnet-mirror.isn-oldenburg.de:8080/dxq-xdp/\crlf
     Msg-To:\ http://metasearch.isn-oldenburg.de/dxq-xqd/\crlf
     \crlf*}
\msg{DXQP-1.0 OK\crlf
     Msg-From:\ http://metasearch.isn-oldenburg.de/dxq-xqd/\crlf
     Msg-To:\ http://physnet-mirror.isn-oldenburg.de:8080/dxq-xdp/\crlf
     \crlf*}
\end{msglist}

\end{document}